\documentclass[pdflatex,sn-mathphys-num]{sn-jnl}

\usepackage[utf8]{inputenc}
\usepackage[english]{babel}
\usepackage[T1]{fontenc}

\usepackage{hyperref}
\usepackage{graphicx}
\usepackage{subcaption}
\usepackage{amsmath,amssymb,amsfonts}
\usepackage{amsthm}
\usepackage{mathrsfs}
\usepackage{xcolor}
\usepackage{booktabs}
\usepackage{algorithm}
\usepackage{algorithmicx}
\usepackage{algpseudocode}
\usepackage{quantikz}
\usepackage{braket}
\usepackage{bm}

\theoremstyle{thmstyleone}

\theoremstyle{thmstyletwo}

\theoremstyle{thmstylethree}

\begin{document}
	
	%%%%%%%%%%%%%%%%%%%%%%%%%%%%%%%%%%%%%%%%%%%%%%
	%% Title
	%%%%%%%%%%%%%%%%%%%%%%%%%%%%%%%%%%%%%%%%%%%%%%
	
	\title[Walsh-Transform Realization of Dense-to-Sparse Quantum State Preparation]
	{Walsh-Transform Realization of Dense-to-Sparse Quantum State Preparation}
	
	%%%%%%%%%%%%%%%%%%%%%%%%%%%%%%%%%%%%%%%%%%%%%%
	%% Authors
	%%%%%%%%%%%%%%%%%%%%%%%%%%%%%%%%%%%%%%%%%%%%%%
	
	\author[1]{\fnm{Emad} \sur{Rezaei Fard Boosari}}
	\email{emad.boosari@gmail.com}	
	
	%%%%%%%%%%%%%%%%%%%%%%%%%%%%%%%%%%%%%%%%%%%%%%
	%% Abstract
	%%%%%%%%%%%%%%%%%%%%%%%%%%%%%%%%%%%%%%%%%%%%%%
	\abstract{
		We propose a hybrid quantum state preparation method based on the Walsh--Hadamard transform within the dense-to-sparse quantum state preparation framework. The method approximately prepares structured classical data in a quantum circuit through an indirect approach.
		Instead of directly preparing the dense amplitude-encoded state, the method first transforms the classical data into the Walsh domain and prepares only the largest-magnitude coefficients using a sparse state-preparation algorithm. 
		The original state can then be approximately recovered through a parallel layer of Hadamard gates.
		Since this reconstruction requires no CNOT gates and has circuit depth 1, the proposed method introduces no additional CNOT or circuit-depth overhead, and its quantum preparation cost is determined by the underlying sparse state-preparation algorithm. Numerical results on representative benchmark signals demonstrate that the proposed method enables accurate state preparation for signals admitting sparse or approximately sparse representations in the Walsh domain.
	}	
	%%%%%%%%%%%%%%%%%%%%%%%%%%%%%%%%%%%%%%%%%%%%%%
	
	
	\maketitle
	
%%%%%%%%%%%%%%%%%%%%%%%%%%%%%%%%%%%%%%%%%%%%%%
%%%%%%%%%%%%%%%%%%%%%%%%%%%%%%%%%%%%%%%%%%%%%%%%%%%%%%%%%%%%%%%%%%%%%%
%%%%%%%%%%%%%%%%%%%%%%%%%%%% Introduction %%%%%%%%%%%%%%%%%%%%%%%%%%%%%
%%%%%%%%%%%%%%%%%%%%%%%%%%%%%%%%%%%%%%%%%%%%%%%%%%%%%%%%%%%%%%%%%%%%%%
\section{Introduction}
Recent advances in quantum computing technologies and applications have substantially strengthened the prospects for realizing practical quantum advantage~\cite{arute2019quantum,daley2022practical,huang2022quantum,kim2023evidence,bluvstein2024logical,google2025quantum}. Achieving such advantages requires advances in both quantum hardware and the implementation of the underlying quantum algorithms. One of the challenging algorithmic primitives that can contribute to practical quantum advantage is quantum state preparation (QSP), which plays an important role in applications including quantum machine learning~\cite{biamonte2017,rebentrost2014}, quantum simulation and quantum chemistry~\cite{low2019hamiltonian,huggins2025efficient}, and quantum algorithms for differential equations and linear systems~\cite{childs2021high,berry2017quantum}. However, since direct preparation of a generic $n$-qubit state requires
$\mathcal{O}(2^n)$ CNOT gates and circuit depth, the cost of QSP can
become a bottleneck that offsets the advantages offered by subsequent
quantum computation.

To address this challenge, numerous approaches have been proposed from different perspectives. 
Variational QSP methods~\cite{zoufal2019quantum,nakaji2022approximate} formulate the loading process as an optimization problem, while tensor-network-based techniques, such as matrix product state (MPS) representations~\cite{melnikov2023quantum, iaconis2024quantum, manabe2025state, ballarin2025efficient}, exploit low-entanglement structures to reduce circuit complexity. 
Moreover, approximate QSP algorithms \cite{moosa2023linear, zylberman2024efficient} demonstrated that allowing a controllable approximation error can substantially decrease the required quantum gates and circuit depth. 

Recently, we introduced dense-to-sparse-QSP (DS-QSP) as an approximate framework for loading structured data into quantum states~\cite{boosari2025hybrid}. DS-QSP employs a hybrid classical--quantum strategy in which a reversible classical transformation first converts the input data into a sparse representation, which is then prepared using a sparse QSP algorithm followed by quantum decompression. Therefore, the performance of DS-QSP depends critically on the choice of the reversible transformation. An effective transformation should simultaneously produce sparse representations for the target signal class while admitting an efficient quantum realization. Identifying transformations that satisfy both requirements is particularly important for practical QSP, where the cost and noise associated with two-qubit gates can become a significant limitation.

One of the unique reversible classical transformations suitable for the DS-QSP framework is the Walsh--Hadamard transform (WHT), which admits a completely CNOT-free quantum implementation with constant circuit depth. 
In classical signal processing, the WHT has been widely used for the representation, analysis, and compression of structured signals, particularly piecewise-constant and binary-valued data~\cite{ahmed2012orthogonal, beauchamp1975walsh, sayood2006introduction}. 
These distinctive properties have recently motivated the development of two probabilistic QSP frameworks~~\cite{zylberman2024efficient, gonzalez2024efficient} based on Walsh transform. The Walsh Series Loader (WSL)~\cite{zylberman2024efficient} is the first QSP algorithm capable of preparing quantum states with a circuit depth and average execution time independent of the number of qubits. 
However, it relies on a repeat-until-success strategy, where improving the approximation accuracy results in a lower success probability.

A different approach based on the WHT was presented by Gonzalez-Conde et al.~\cite{gonzalez2024efficient}. 
Their method exploits the Walsh transformation as an efficient state-preparation primitive for polynomial transformations using quantum singular value transformation (QSVT)~\cite{low2017optimal, gilyen2019quantum}. Their approach demonstrates the utility of Walsh sparsity for efficient function encoding, while the preparation of the resulting polynomial states involves amplitude amplification. In contrast to a general QSP framework for arbitrary structured signals, their construction is specifically developed for polynomial transformations through QSVT.

In this work, we present a Walsh-transform realization of the DS-QSP framework for the approximate preparation of one-dimensional classical signals. 
The proposed paradigm provides a straightforward approach for structured classical signals that possess sparse or compressible representations in the Walsh domain. 
In contrast to the WSL and QSVT-based approaches, which employ probabilistic preparation procedures and may require amplitude
amplification to increase their success probability, the proposed framework provides a deterministic preparation procedure that can be
implemented using either ancilla-free or ancilla-assisted sparse QSP.

According to the DS-QSP workflow, the WHT is employed in the classical pre-processing stage to compress the information of the input signal in the Walsh domain. 
When the resulting representation is not sufficiently sparse, a hard-thresholding procedure retains only the $d$ largest-magnitude coefficients to obtain a sparse representation. 
Since thresholding discards part of the information that cannot be recovered during quantum decompression, the resulting quantum state provides an approximation to the target state. 
The quantum stage then consists of sparse QSP~\cite{li2024nearly, farias2025quantum, zhang2022quantum} for preparing the retained Walsh coefficients, followed by quantum decompression using the inverse WHT. Since the inverse quantum WHT is realized by the tensor product of Hadamard gates, it has circuit depth $\mathcal{O}(1)$ and introduces no CNOT overhead. Consequently, the quantum resource requirements of the proposed realization are determined primarily by the sparse state-preparation stage and the Walsh-domain compressibility of the input signal.

To characterize the properties and practical effectiveness of the proposed Walsh-based realization of DS-QSP, we perform a systematic numerical investigation using representative one-dimensional signals. We first examine the Walsh-domain compressibility of signals with different structural characteristics and its effect on the required number of retained coefficients. We then investigate the resulting reconstruction accuracy as a function of the Walsh sparsity, followed by an analysis of how the required sparsity scales with the signal dimension and the prescribed approximation accuracy. These experiments allow us to identify the signal classes for which the Walsh-based realization can provide substantial reductions in sparse QSP resources and to characterize the conditions under which its compression advantage is preserved.

The remainder of this paper is organized as follows. 
Section~\ref{sec:ds-qsp} introduces the DS-QSP framework and the corresponding approximation and quantum-state reconstruction measures. Section~\ref{sec:method} presents the proposed Walsh--Hadamard realization of DS-QSP, including its quantum resource requirements. 
Section~\ref{sec:examples} investigates the Walsh-domain compressibility and reconstruction performance for representative benchmark signals. Finally, Section~\ref{sec:conclusion} concludes the paper.

%%%%%%%%%%%%%%%%%%%%%%%%%%%%%%%%%%%%%%%%%%%%%%%%%%%%%%%%%%%%%
%%%%%%%%%%%%%%%%%%%%%%%%%%%%%%%%%%%%%%%%%%%%%%%%%%%%%%%%%%%%%
%%%%%%%%%%%%%%%%%%%% DS-QSP Framework %%%%%%%%%%%%%%%%%%%%%%%
%%%%%%%%%%%%%%%%%%%%%%%%%%%%%%%%%%%%%%%%%%%%%%%%%%%%%%%%%%%%%
\section{DS-QSP Framework}\label{sec:ds-qsp}

DS-QSP is a hybrid classical--quantum framework that reduces the direct preparation cost of a dense quantum state to sparse quantum state preparation followed by quantum decompression~\cite{boosari2025hybrid}. 
The framework can naturally be generalized to higher-dimensional data, such as images~\cite{boosari2026hybrid}.
The central idea is to exploit a reversible classical transformation that converts the input data into a sparse or approximately sparse
representation.

Consider a classical vector
%---------------------------------------
\begin{equation}\label{eq:classical_vector}
	\bm{x} = [x_0,x_1,\ldots,x_{N-1}]^T, \qquad N=2^n,
\end{equation}
%---------------------------------------
whose target amplitude-encoded quantum state is
%---------------------------------------
\begin{equation}\label{eq:amplitude_encoding}
	\ket{\bm{x}} = \frac{1}{\|\bm{x}\|_2} \sum_{j=0}^{N-1}x_j\ket{j}.
\end{equation}
%---------------------------------------
Direct preparation of a generic $n$-qubit state requires an exponential number of elementary gates, including
$\mathcal{O}(2^n)$ CNOT gates~\cite{mottonen2004quantum,iten2016}.
DS-QSP addresses this cost by separating the preparation into a classical compression stage and a quantum decompression stage.

%@@@@@@@@@@@@@@@@@@@@@@@@@@@@@@@@@@@@@@@@@@@@@@@@@@@@@@@@@@@@
\begin{figure}[t]
	\centering	
	\begin{subfigure}[t]{0.30\textwidth}
		\centering
		\includegraphics[width=\textwidth]{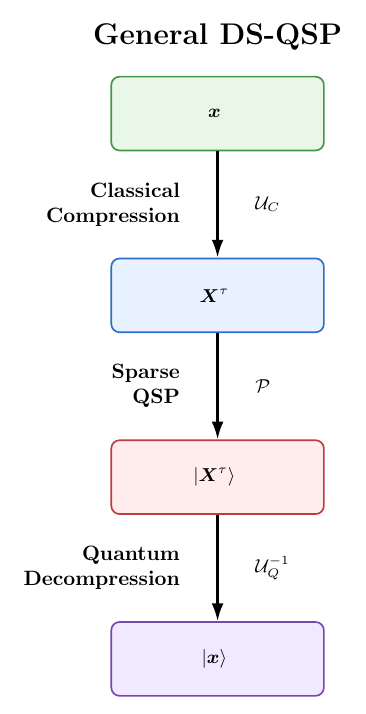}
		\caption{}
		\label{fig:workflow}
	\end{subfigure}
	\hspace{0.07\textwidth}
	\begin{subfigure}[t]{0.375\textwidth}
		\centering
		\includegraphics[width=\textwidth]{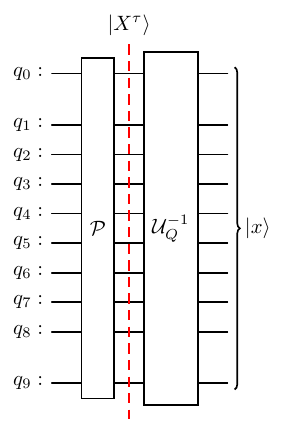}
		\caption{}
		\label{fig:quantum_circuit}
	\end{subfigure}	
	\caption{
		(a) General workflow of the DS-QSP framework.
		(b) Quantum circuit corresponding to the quantum stage of the
		framework.
	}
	\label{fig:method}
\end{figure}
%@@@@@@@@@@@@@@@@@@@@@@@@@@@@@@@@@@@@@@@@@@@@@@@@@@@@@@@@@@@@

The DS-QSP proposes an indirect preparation workflow which has been illustrated in Fig.~\ref{fig:workflow}. In the classical stage, the input vector is first transformed according to
%---------------------------------------
\begin{equation}\label{eq:classical_transform}
	\bm{X} = \mathcal{U}_C\bm{x} = [X_0,X_1,\ldots,X_{N-1}]^T,
\end{equation}
%---------------------------------------
where $\mathcal{U}_C$ denotes a reversible classical transformation. 
If the transformed vector is already sparse, it can be directly used as the output of the classical stage. 
Otherwise, a coefficient-reduction procedure, such as top-$d$ thresholding, retains the $d$ largest-magnitude coefficients and sets the remaining coefficients to zero, producing the sparse representation $\bm{X}^{\tau}$.

The resulting vector $\bm{X}^{\tau}$ provides either an exact or an approximate representation of the transformed data, depending on
whether coefficient reduction is required. To quantify the information discarded by the coefficient-reduction procedure, we define the
relative transform-domain approximation error as
%---------------------------------------
\begin{equation}\label{eq:relative_error}
	\epsilon = \frac{\|\bm{X}-\bm{X}^{\tau}\|_2} {\|\bm{X}\|_2}.
\end{equation}
%---------------------------------------
For a prescribed tolerance $\epsilon$, the coefficient-reduction procedure retains the minimum number of coefficients required to
satisfy Eq.~\eqref{eq:relative_error}. In the absence of coefficient reduction, $\bm{X}^{\tau}=\bm{X}$ and $\epsilon=0$.

The resulting $d$-sparse vector is then prepared as a quantum state using a sparse QSP algorithm. Denoting the corresponding
state-preparation operator by $\mathcal{P}$, the quantum state after the sparse preparation stage is
%---------------------------------------
\begin{equation}\label{eq:sparse_state_preparation}
	\ket{\bm{X}} = \mathcal{P}\ket{0}^{\otimes n} = \frac{1}{\|\bm{X}^{\tau}\|_2} \sum_{j=0}^{N-1}X_j^\tau\ket{j}.
\end{equation}
%---------------------------------------

The DS-QSP framework does not require a particular sparse encoder. 
Different sparse QSP algorithms can therefore be incorporated according to the desired trade-off between circuit depth, gate count,
and ancillary-qubit requirements. In the ancilla-free setting, sparse state preparation can be performed with asymptotic circuit size
$\mathcal{O}(nd)$, while ancillary-assisted constructions can reduce the circuit size~\cite{li2024nearly} to
%---------------------------------------
\begin{equation}\label{eq:sqsp_ancilla}
	\mathcal{O}\left(
	\frac{nd}{\log(n+m)}+n
	\right),
\end{equation}
%---------------------------------------
where $m$ denotes the number of ancillary qubits.

Finally, the sparse transform-domain state is mapped back to the original data domain through a quantum implementation of the inverse
classical transformation,
%---------------------------------------
\begin{equation}\label{eq:quantum_decompression}
	\ket{\tilde{\bm{x}}} = \mathcal{U}_Q^{-1}\ket{\bm{X}},
\end{equation}
%---------------------------------------
where $\mathcal{U}_Q^{-1}$ denotes the quantum realization of the inverse of $\mathcal{U}_C$. 
The difference between the prepared state $\ket{\tilde{\bm{x}}}$ and target state $\ket{\bm{x}}$ can be evaluated using the trace distance~\cite{nielsen2010quantum}.

Figure~\ref{fig:quantum_circuit} represents the quantum circuit of the DS-QSP method in two blocks: (i) sparse encoder and (ii) quantum inverse transformation. The total CNOT gate cost for the DS-QSP scheme can therefore be written as
%---------------------------------------
\begin{equation}\label{eq:dsqsp_complexity}
	C_{\mathrm{DS-QSP}} = C_{\mathrm{sparse}} + C_{\mathcal{U}_Q},
\end{equation}
%---------------------------------------
where $C_{\mathrm{sparse}}$ denotes the CNOT cost of preparing the $d$-sparse transform-domain state $\bm{X}^{\tau}$, and
$C_{\mathcal{U}_Q}$ denotes the CNOT cost of implementing the quantum inverse transformation $\mathcal{U}_Q^{-1}$.

For an effective DS-QSP realization, two complementary requirements should be satisfied: the classical transformation should produce a
sufficiently sparse representation of the target data, and its quantum inverse transformation should admit an efficient implementation.

In the following section, we specialize the DS-QSP framework to the WHT and show that its inverse transformation can be implemented solely with single-qubit Hadamard gates, resulting in a CNOT-free quantum decompression stage with constant circuit depth.

%%%%%%%%%%%%%%%%%%%%%%%%%%%%%%%%%%%%%%%%%%%%%%%%%%%%%%%%%%%%%
%%%%%%%%%%%%%%%%%%%%%%%%%%%% Method %%%%%%%%%%%%%%%%%%%%%%%%%%
%%%%%%%%%%%%%%%%%%%%%%%%%%%%%%%%%%%%%%%%%%%%%%%%%%%%%%%%%%%%%
\section{Walsh Realization of DS-QSP}\label{sec:method}

This section presents a WHT realization of the DS-QSP framework and analyzes the properties that distinguish it from other transform-based realizations. 
Following the classical preprocessing step in Eq.~\eqref{eq:classical_transform}, the reversible classical transformation is chosen as the normalized WHT,
%-----------------------------------------
\begin{equation}
	\mathcal U_C = \frac{1}{\sqrt N}W_N,
\end{equation}
%-----------------------------------------
where $W_N\in\{\pm1\}^{N\times N}$ denotes the Walsh--Hadamard matrix of order $N$. Throughout this work, the WHT is generated recursively from the Hadamard matrices introduced by Sylvester~\cite{sylvester1867lx},
%-----------------------------------------
\begin{equation}
	W_1=[1],\qquad
	W_2=
	\begin{bmatrix}
		1 & 1\\
		1 & -1
	\end{bmatrix},
\end{equation}
%-----------------------------------------
and
%-----------------------------------------
\begin{equation}
	W_{2^k} = 
	\begin{bmatrix}
		W_{2^{k-1}} & W_{2^{k-1}}\\
		W_{2^{k-1}} & -W_{2^{k-1}}
	\end{bmatrix}= W_2\otimes W_{2^{k-1}}, \qquad k\ge 2 \in \mathbb{N},
\end{equation}
%-----------------------------------------
where $\otimes$ denotes the Kronecker product. Since
%-----------------------------------------
\begin{equation}
	H=\frac{1}{\sqrt2}W_2,
\end{equation}
%-----------------------------------------
the normalized $2\times2$ Walsh--Hadamard matrix coincides exactly with the single-qubit Hadamard gate, providing a direct connection between the classical transform and its quantum realization.
Thus, the transformed coefficient vector is obtained as
%-----------------------------------------
\begin{equation}
	\bm X = \mathcal U_C\bm x = \frac{1}{\sqrt N}W_N\bm x.
\end{equation}
%-----------------------------------------
Since the normalized Walsh--Hadamard matrix is symmetric and orthogonal,
%-----------------------------------------
\begin{equation}
	\mathcal U_C^{-1} = \mathcal U_C^{T} = \mathcal U_C,
\end{equation}
%-----------------------------------------
the inverse transformation is identical to the forward transform.

The distinguishing feature of the proposed realization appears in the quantum decompression stage. While Eq.~\eqref{eq:quantum_decompression} represents a general inverse transformation within the DS-QSP framework, the inverse quantum WHT is implemented exactly as
%---------------------------------------
\begin{equation}
	U_{\mathrm{WHT}} = H^{\otimes n},
\end{equation}
%---------------------------------------
which $U_{\mathrm{WHT}}^{-1} = U_{\mathrm{WHT}}$ is its self-inverse.

The target quantum state is reconstructed as
%---------------------------------------
\begin{equation}
	\ket{\bm{\tilde{x}}} \approx H^{\otimes n}\ket{\bm X}.
\end{equation}
%---------------------------------------
In contrast to DS-QSP realizations based on the quantum Fourier transform and the quantum packet Haar wavelet transform, whose quantum decompression circuits require $\mathcal{O}(n^2)$ CNOT gates and circuit depth $\mathcal{O}(n)$, the inverse WHT is implemented in quantum circuit exactly by a single parallel layer of Hadamard gates.
Consequently, the quantum decompression stage is completely free of CNOT gates and contributes only a depth-one overhead to the overall state-preparation circuit. Thus, from Eq.~\eqref{eq:dsqsp_complexity},
%---------------------------------------
\begin{equation}
	C_{\mathrm{Walsh}} = C_{\mathrm{sparse}}.
\end{equation}
%---------------------------------------
The resulting circuit depth is therefore determined asymptotically by that of the selected sparse state-preparation algorithm.

%%%%%%%%%%%%%%%%%%%%%%%%%%%%%%%%%%%%%%%%%%%%%%%%%%%%%%%%%%%%%
%%%%%%%%%%%%%%%%%%%%%%%%%%%%%%%%%%%%%%%%%%%%%%%%%%%%%%%%%%%%%
%%%%%%%%%%%%%%%%%%%% Numerical Results %%%%%%%%%%%%%%%%%%%%%%
%%%%%%%%%%%%%%%%%%%%%%%%%%%%%%%%%%%%%%%%%%%%%%%%%%%%%%%%%%%%%
\section{Numerical Results}\label{sec:examples}

The quantum resource requirements of the proposed Walsh-based DS-QSP realization are primarily governed by the Walsh-domain sparsity $d$. 
We therefore investigate the Walsh-domain compressibility of representative benchmark signals and characterize how the required
sparsity $d$ depends on the signal structure, signal dimension, and prescribed approximation accuracy. 

%%%%%%%%%%%%%%%%%%%%%%%%%%%%%%%%%%%%%%%%%%%%%%%%%%%%%%%%%%%%%
%%%%%%%%%%%%%%%%%%%% Signal Classes and Compressibility %%%%%
%%%%%%%%%%%%%%%%%%%%%%%%%%%%%%%%%%%%%%%%%%%%%%%%%%%%%%%%%%%%%

\begin{figure}[t]
	\centering
	\includegraphics[width=0.95\linewidth]{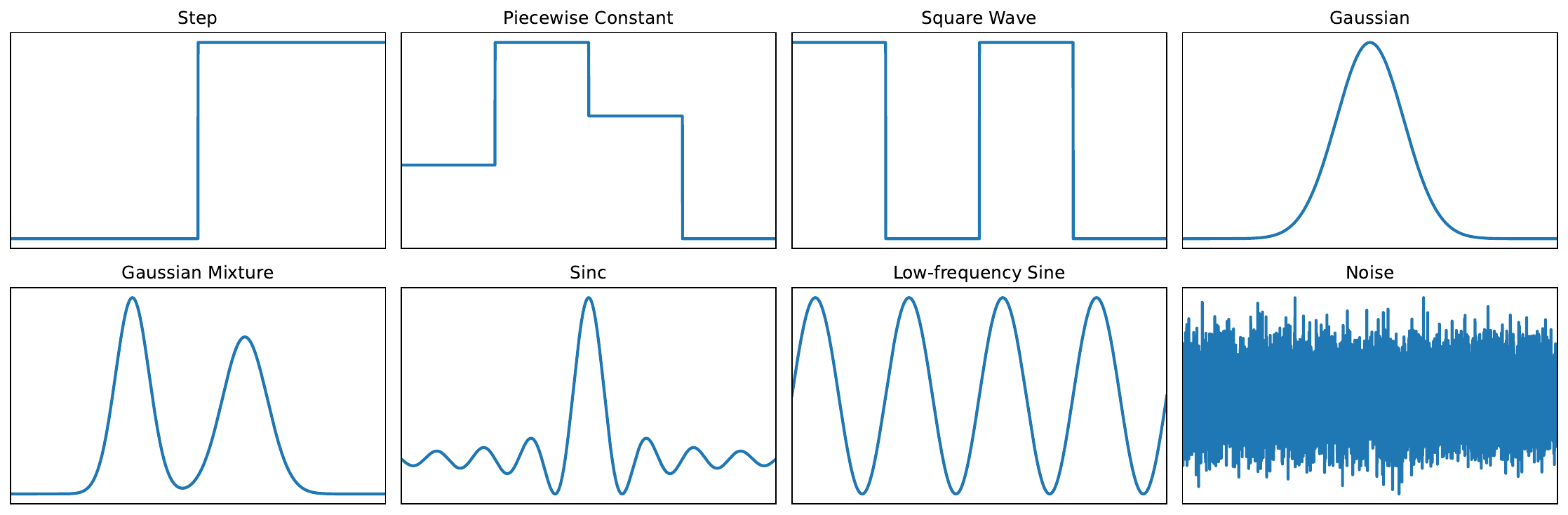}
	\caption{
		Representative one-dimensional benchmark signals used in the numerical experiments, spanning a range of structural
		characteristics from highly structured and Walsh-sparse signals to unstructured white noise.
	}
	\label{fig:benchmark_signals}
\end{figure}

We first consider eight representative benchmark signals spanning a range of structural characteristics. These signals provide examples ranging from exactly or highly Walsh-sparse structured signals to essentially incompressible unstructured data. This range allows the dependence of Walsh-domain compressibility on the underlying signal structure to be examined systematically. 
Figure~\ref{fig:benchmark_signals} summarizes the benchmark signals considered throughout the numerical experiments.

%%%%%%%%%%%%%%%%%%%%%%%%%%%%%%%%%%%%%%%%%%%%%%%%%%%%%%%%%%%%%
%%%%%%%%%%%%%%%%%%%% Walsh Compressibility %%%%%%%%%%%%%%%%%%
%%%%%%%%%%%%%%%%%%%%%%%%%%%%%%%%%%%%%%%%%%%%%%%%%%%%%%%%%%%%%

\begin{figure}[t]
	\centering
	
	\begin{subfigure}[b]{0.48\linewidth}
		\centering
		\includegraphics[width=\linewidth]{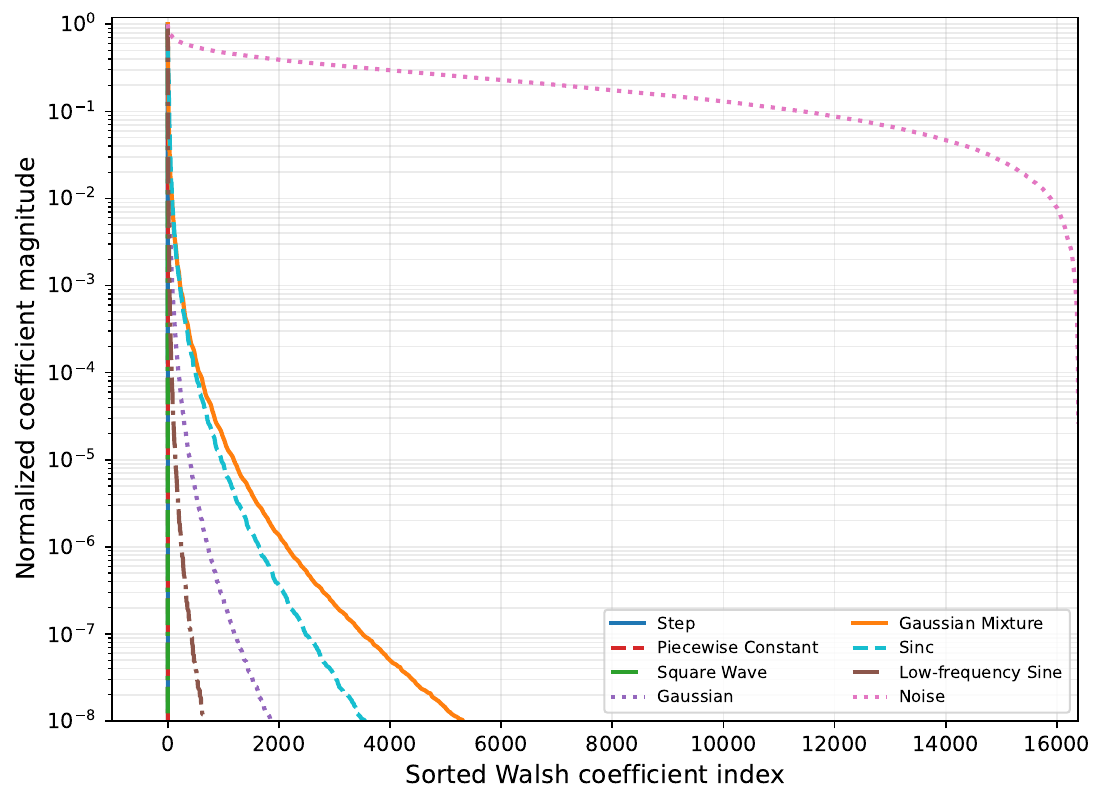}
		\caption{}
		\label{fig:walsh_coeffs}
	\end{subfigure}
	\hfill
	\begin{subfigure}[b]{0.48\linewidth}
		\centering
		\includegraphics[width=\linewidth]{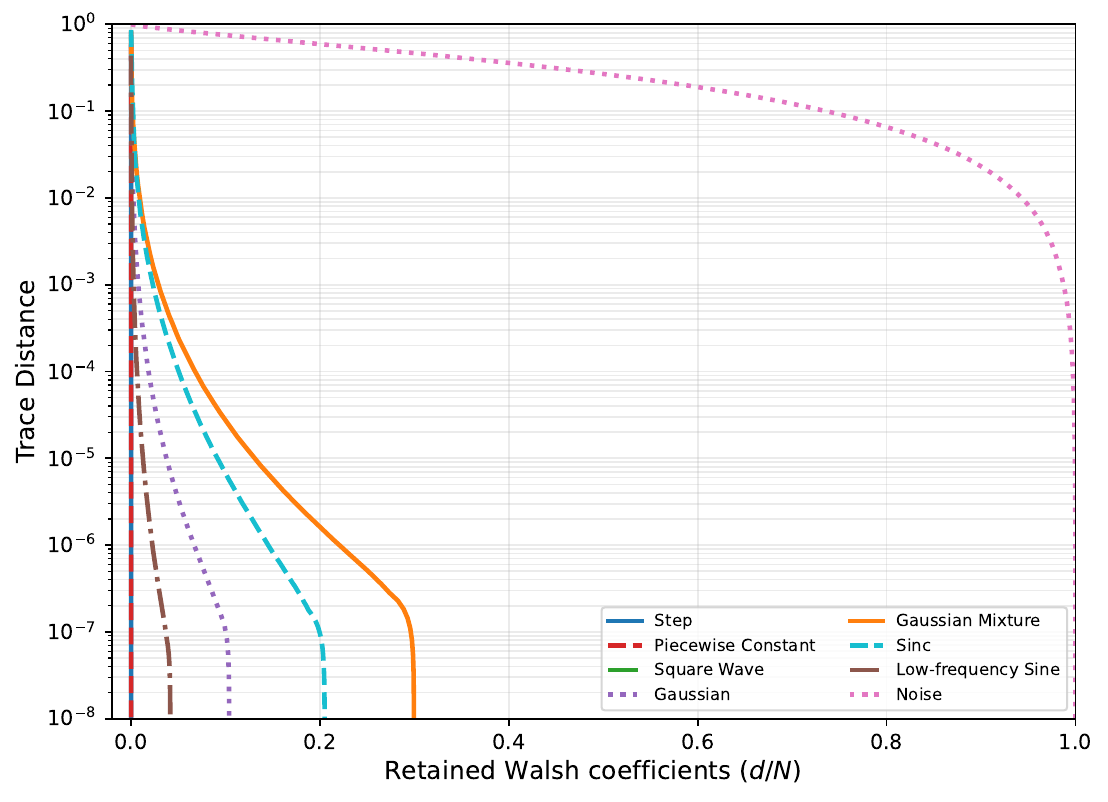}
		\caption{}
		\label{fig:trace_distance}
	\end{subfigure}	
	\caption{
		Walsh-domain compressibility and quantum-state reconstruction for
		the representative benchmark signals.
		(a) Sorted Walsh coefficient magnitudes for the eight benchmark
		signals.
		(b) Trace distance between the target and reconstructed quantum
		states as a function of the normalized number of retained Walsh coefficients $d/N$.
	}               
	\label{fig:walsh_reconstruction}
\end{figure}

Figure~\ref{fig:walsh_reconstruction} summarizes the Walsh-domain compressibility and the resulting quantum-state reconstruction
performance for the eight benchmark signals. The results can be classified into three broad groups according to their Walsh-domain
structure. The first group consists of exactly Walsh-sparse signals, including the step, piecewise-constant, and square-wave signals, whose representations contain only two, three, and one non-zero coefficients, respectively. The second group consists of approximately Walsh-sparse signals, including the low-frequency sinusoid, Gaussian, Gaussian-mixture, and sinc signals, whose Walsh coefficients are not exactly sparse but exhibit sufficient concentration to permit high fidelity reconstruction from a small subset of coefficients. The final group is represented by white noise, whose Walsh coefficients remain broadly distributed and therefore provide little opportunity for compression.
%@@@@@@@@@@@@@@@@@@@@@@@@@@@@@@@@@@@@@@@@@@@@@@@@@@@@@@@@@@@@@@
%@@@@@@@@@@@@@@@@@@@@@@@@@@@@@@@@@@@@@@@@@@@@@@@@@@@@@@@@@@@@@@

%@@@@@@@@@@@@@@@@@@@@@@@@@@@@@@@@@@@@@@@@@@@@@@@@@@@@@@@@@@@@@@
%@@@@@@@@@@@@@@@@@@@@@@@@@@@@@@@@@@@@@@@@@@@@@@@@@@@@@@@@@@@@@@

%%%%%%%%%%%%%%%%%%%%%%%%%%%%%%%%%%%%%%%%%%%%%%%%%%%%%%%%%%%%%
%%%%%%%%%%%%%%%%%%%% Reconstruction Performance %%%%%%%%%%%%%
%%%%%%%%%%%%%%%%%%%%%%%%%%%%%%%%%%%%%%%%%%%%%%%%%%%%%%%%%%%%%

Having established the three classes of Walsh-domain structure, we next examine their implications for quantum-state reconstruction.
Figure~\ref{fig:trace_distance} shows the trace distance between the target and reconstructed states as a function of the retained Walsh coefficients. The exactly Walsh-sparse signals achieve exact reconstruction with only a constant number of coefficients, while the approximately Walsh-sparse signals converge to the target state using substantially fewer coefficients than the full signal dimension. In contrast, the Walsh-incompressible white noise requires nearly the full Walsh representation to achieve a comparable reconstruction accuracy. These results demonstrate that the reduction in quantum state-preparation resources is substantial for exactly and approximately Walsh-sparse signals, whereas incompressible signals offer little advantage over direct dense state preparation.

\begin{figure}[t]
	\centering
	
	\begin{subfigure}[b]{0.48\linewidth}
		\centering
		\includegraphics[width=\linewidth]{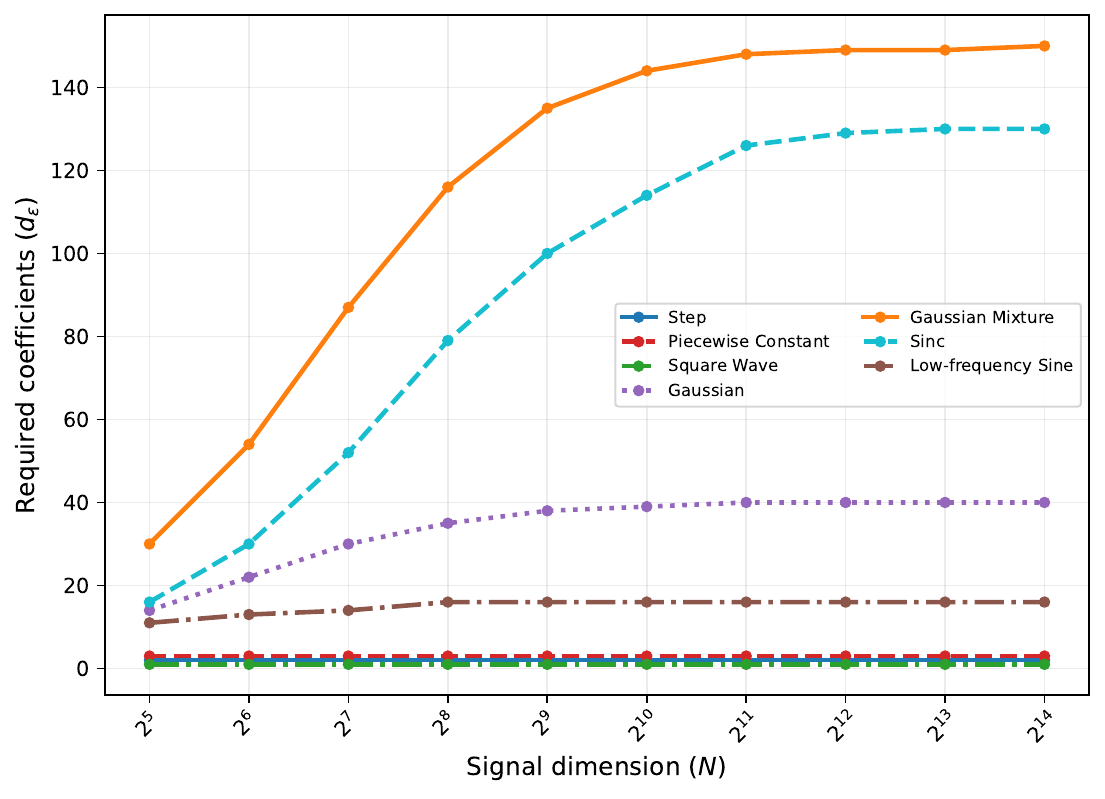}
		\caption{}
		\label{fig:scaling_dim}
	\end{subfigure}
	\hfill
	\begin{subfigure}[b]{0.48\linewidth}
		\centering
		\includegraphics[width=\linewidth]{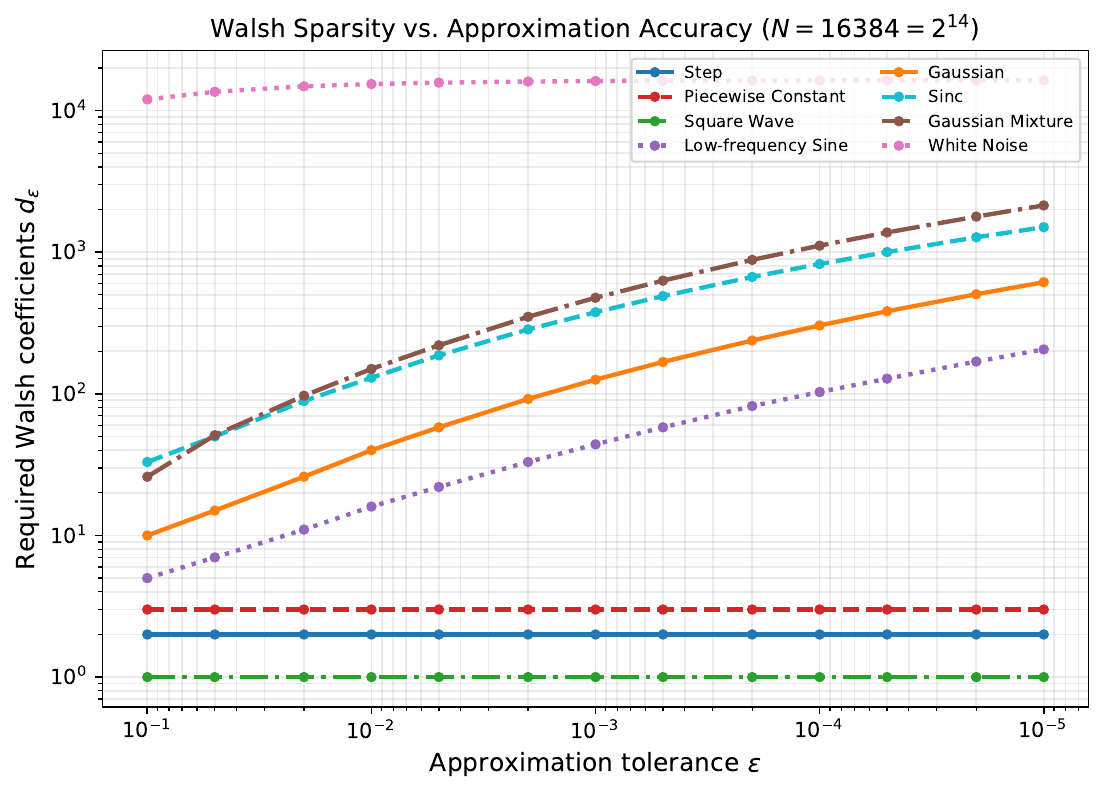}
		\caption{}
		\label{fig:scaling_accuracy}
	\end{subfigure}
	
	\caption{Scaling behavior of the proposed Walsh-based DS-QSP framework. 
		(a) Minimum number of retained Walsh coefficients required to satisfy a fixed relative approximation tolerance of
		$\epsilon=10^{-2}$ as the signal dimension increases from $N=2^5$ to $N=2^{14}$. 		
		(b) Minimum number of retained Walsh coefficients $d_\epsilon$ required to achieve different relative
		approximation tolerances $\epsilon$ for fixed signal dimension $N=2^{14}$. The exactly Walsh-sparse signals remain at
		$1$, $2$, and $3$ coefficients for the square-wave, step, and piecewise-constant signals, respectively. The low-frequency
		sinusoid increases from $5$ to $206$ coefficients as $\epsilon$ decreases from $10^{-1}$ to $10^{-5}$. 
		At $\epsilon=10^{-5}$, the Gaussian, sinc, and Gaussian-mixture signals require $614$, $1502$, and $2141$ coefficients,
		corresponding to $3.75\%$, $9.17\%$, and $13.07\%$ of the available coefficients, respectively, while white noise
		requires $16375$ coefficients ($99.95\%$ of $N$).
	}
	\label{fig:scaling}
\end{figure}

%%%%%%%%%%%%%%%%%%%%%%%%%%%%%%%%%%%%%%%%%%%%%%%%%%%%%%%%%%%%%
%%%%%%%%%%%%%%%%%%%% Asymptotic Scaling %%%%%%%%%%%%%%%%%%%%%
%%%%%%%%%%%%%%%%%%%%%%%%%%%%%%%%%%%%%%%%%%%%%%%%%%%%%%%%%%%%%
Finally, we investigate how the required Walsh sparsity scales with the signal dimension and the prescribed approximation accuracy.
The dimension scaling in Fig.~\ref{fig:scaling_dim} shows that the
exactly Walsh-sparse signals require a constant number of coefficients
as the signal dimension increases, while the approximately
Walsh-sparse signals exhibit substantially slower growth than the
signal dimension. The Walsh-incompressible white-noise signal is
excluded from this plot because its near-linear scaling obscures the
behavior of the structured signals. These results indicate that, for Walsh-sparse and
Walsh-compressible signals, the required sparsity is determined primarily by the signal structure rather than by the ambient dimension, whereas incompressible signals retain a resource requirement that scales with the full signal dimension.

Figure~\ref{fig:scaling_accuracy} examines how the required Walsh sparsity changes with the prescribed approximation accuracy for the
fixed signal dimension $N=16384$ ($n=14$). Exactly Walsh-sparse signals retain the same number of coefficients independently of the prescribed accuracy. Approximately Walsh-sparse signals require progressively more coefficients as the accuracy is tightened, but remain substantially sparser than the full representation over the investigated range. In contrast, Walsh-incompressible signals rapidly approach the full signal dimension as higher accuracy is required. These results show that the benefit of Walsh-domain compression depends not only on the signal structure but also on the required approximation accuracy.

%@@@@@@@@@@@@@@@@@@@@@@@@@@@@@@@@@@@@@@@@@@@@@@@@@@@@@@@@@@@@@@
%@@@@@@@@@@@@@@@@@@@@@@@@@@@@@@@@@@@@@@@@@@@@@@@@@@@@@@@@@@@@@@

%%%%%%%%%%%%%%%%%%%%%%%%%%%%%%%%%%%%%%%%%%%%%%%%%%%%%%%%%%%%%
%%%%%%%%%%%%%%%%%%%% Reconstruction Visualization %%%%%%%%%%%
%%%%%%%%%%%%%%%%%%%%%%%%%%%%%%%%%%%%%%%%%%%%%%%%%%%%%%%%%%%%%

\begin{figure}[t]
	\centering
	\includegraphics[width=0.95\linewidth]{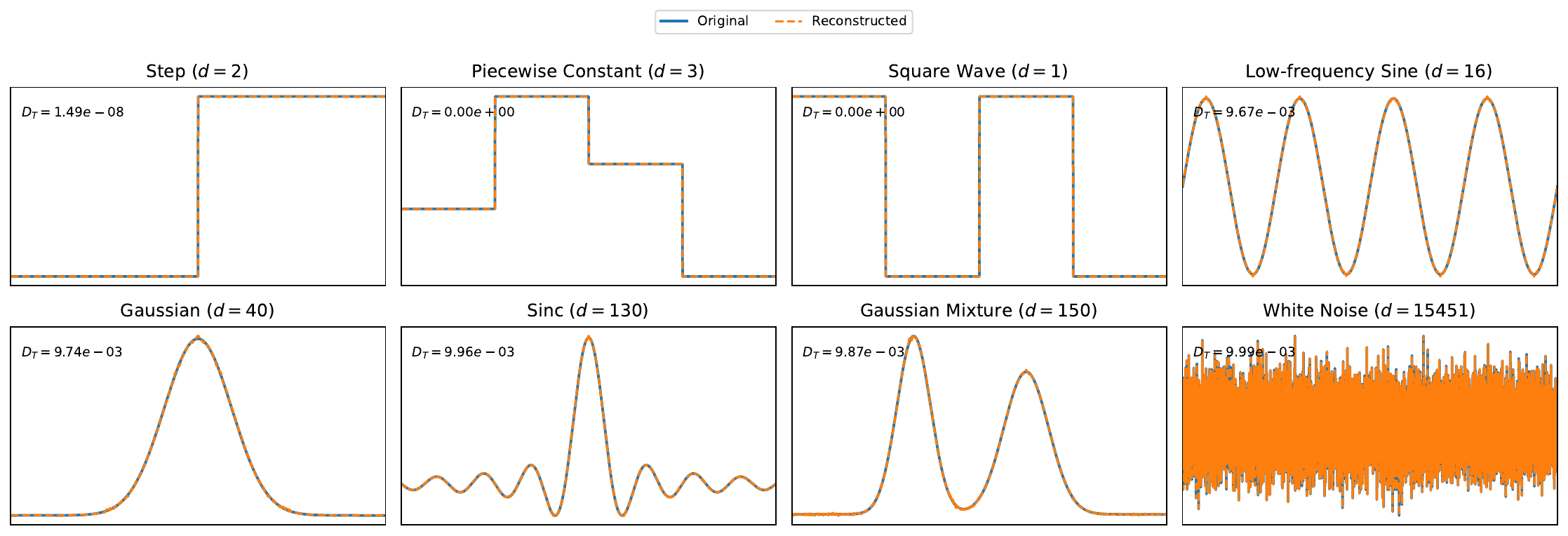}
	\caption{
		Original and Walsh-domain reconstructed benchmark signals for
		$N=16384$ and relative approximation tolerance
		$\epsilon=10^{-2}$. For each signal, the minimum number
		$d_\epsilon$ of largest-magnitude Walsh coefficients required to
		satisfy the prescribed tolerance is retained before applying the
		inverse WHT.
	}
	\label{fig:reconstruction_signals}
\end{figure}

To complement the quantitative reconstruction analysis, we finally visualize the reconstructed signals obtained after Walsh-domain
thresholding. Figure~\ref{fig:reconstruction_signals} compares the original and reconstructed signals for all eight benchmark classes at $N=16384$ and a fixed relative approximation tolerance of $\epsilon=10^{-2}$. For each signal, the minimum number of Walsh
coefficients satisfying this tolerance is retained before applying the inverse WHT. The resulting reconstructions provide a direct visual illustration of the compression behavior observed in the previous experiments. The exactly or highly Walsh-sparse signals are
reconstructed using only a very small number of coefficients, while the Gaussian, sinc, and Gaussian-mixture signals retain progressively more Walsh components. In contrast, the reconstruction of white noise requires almost the full Walsh representation, consistent with its lack of Walsh-domain compressibility.

%@@@@@@@@@@@@@@@@@@@@@@@@@@@@@@@@@@@@@@@@@@@@@@@@@@@@@@@@@@@@@@
%@@@@@@@@@@@@@@@@@@@@@@@@@@@@@@@@@@@@@@@@@@@@@@@@@@@@@@@@@@@@@@

%%%%%%%%%%%%%%%%%%%%%%%%%%%%%%%%%%%%%%%%%%%%%%%%%%%%%%%%%%%%%
%%%%%%%%%%%%%%%%%%%% Overall Observation %%%%%%%%%%%%%%%%%%%%
%%%%%%%%%%%%%%%%%%%%%%%%%%%%%%%%%%%%%%%%%%%%%%%%%%%%%%%%%%%%%

Overall, the numerical results demonstrate that the WHT, together with its CNOT-free quantum implementation, provides
an effective realization of DS-QSP for structured signals admitting
sparse or approximately sparse Walsh representations. The required
quantum resources are therefore determined primarily by the
Walsh-domain compressibility of the input signal.

%%%%%%%%%%%%%%%%%%%%%%%%%%%%%%%%%%%%%%%%%%%%%%%%%%%%%%%%%%%%%
%%%%%%%%%%%%%%%%%%%%%%%%%%%%%%%%%%%%%%%%%%%%%%%%%%%%%%%%%%%%%
%%%%%%%%%%%%%%%%%%%%%%%% Conclusion %%%%%%%%%%%%%%%%%%%%%%%%%
%%%%%%%%%%%%%%%%%%%%%%%%%%%%%%%%%%%%%%%%%%%%%%%%%%%%%%%%%%%%%

\section{Conclusion}\label{sec:conclusion}

In this work, we presented a WHT realization of the DS-QSP framework for the approximate preparation of classical signals. The proposed approach uses the WHT as a reversible classical compression stage, followed by sparse QSP of the retained Walsh coefficients and quantum decompression through the inverse WHT. Since the quantum inverse of WHT is implemented solely by the tensor product of Hadamard gates, the decompression stage is CNOT-free and has constant circuit depth.
Consequently, the quantum resource requirements of the proposed realization are primarily determined by the Walsh-domain sparsity and
the complexity of the selected sparse state-preparation algorithm.

The numerical results demonstrate that this approach provides an effective route for preparing structured signals that admit sparse or
approximately sparse Walsh representations. The results also show that the benefit of Walsh-based DS-QSP is inherently dependent on the
Walsh-domain compressibility of the target signal. The modular nature of the framework allows improved sparse state-preparation algorithms to be incorporated without changing the Walsh compression or decompression stages. An immediate direction for future work is the extension of the proposed approach to higher-dimensional structured data, including quantum image preparation using multidimensional Walsh transforms.
	
\section*{Funding}
This research did not receive any specific grant from funding agencies in the public, commercial, or not-for-profit sectors.

%%%%%%%%%%%%%%%%%%%%%%%%%%%%%%%%%%%%%%%%%%%%%%
\bibliography{sn-bibliography}

@article{biamonte2017,
  title={Quantum machine learning},
  author={Biamonte, Jacob and Wittek, Peter and Pancotti, Nicola and Rebentrost, Patrick and Wiebe, Nathan and Lloyd, Seth},
  journal={Nature},
  volume={549},
  number={7671},
  pages={195--202},
  year={2017},
  publisher={Nature Publishing Group UK London}
}

@article{rebentrost2014,
  title={Quantum support vector machine for big data classification},
  author={Rebentrost, Patrick and Mohseni, Masoud and Lloyd, Seth},
  journal={Physical review letters},
  volume={113},
  number={13},
  pages={130503},
  year={2014},
  publisher={APS}
}

@article{low2019hamiltonian,
  title={Hamiltonian simulation by qubitization},
  author={Low, Guang Hao and Chuang, Isaac L},
  journal={Quantum},
  volume={3},
  pages={163},
  year={2019},
  publisher={Verein zur F{\"o}rderung des Open Access Publizierens in den Quantenwissenschaften}
}

@article{berry2017quantum,
  title={Quantum algorithm for linear differential equations with exponentially improved dependence on precision},
  author={Berry, Dominic W and Childs, Andrew M and Ostrander, Aaron and Wang, Guoming},
  journal={Communications in Mathematical Physics},
  volume={356},
  number={3},
  pages={1057--1081},
  year={2017},
  publisher={Springer}
}

@article{iten2016,
  title={Quantum circuits for isometries},
  author={Iten, Raban and Colbeck, Roger and Kukuljan, Ivan and Home, Jonathan and Christandl, Matthias},
  journal={Physical Review A},
  volume={93},
  number={3},
  pages={032318},
  year={2016},
  publisher={APS}
}

@article{nakaji2022approximate,
  title={Approximate amplitude encoding in shallow parameterized quantum circuits and its application to financial market indicators},
  author={Nakaji, Kouhei and Uno, Shumpei and Suzuki, Yohichi and Raymond, Rudy and Onodera, Tamiya and Tanaka, Tomoki and Tezuka, Hiroyuki and Mitsuda, Naoki and Yamamoto, Naoki},
  journal={Physical Review Research},
  volume={4},
  number={2},
  pages={023136},
  year={2022},
  publisher={APS}
}

@article{zoufal2019quantum,
  title={Quantum generative adversarial networks for learning and loading random distributions},
  author={Zoufal, Christa and Lucchi, Aur{\'e}lien and Woerner, Stefan},
  journal={npj Quantum Information},
  volume={5},
  number={1},
  pages={103},
  year={2019},
  publisher={Nature Publishing Group UK London}
}

@article{zhang2022quantum,
  title={Quantum state preparation with optimal circuit depth: Implementations and applications},
  author={Zhang, Xiao-Ming and Li, Tongyang and Yuan, Xiao},
  journal={Physical Review Letters},
  volume={129},
  number={23},
  pages={230504},
  year={2022},
  publisher={APS}
}

@article{li2024nearly,
  title={Nearly optimal circuit size for sparse quantum state preparation},
  author={Li, Lvzhou and Luo, Jingquan},
  journal={arXiv preprint arXiv:2406.16142},
  year={2024}
}

@article{moosa2023linear,
  title={Linear-depth quantum circuits for loading Fourier approximations of arbitrary functions},
  author={Moosa, Mudassir and Watts, Thomas W and Chen, Yiyou and Sarma, Abhijat and McMahon, Peter L},
  journal={Quantum Science and Technology},
  volume={9},
  number={1},
  pages={015002},
  year={2023},
  publisher={IOP Publishing}
}

@article{gonzalez2024efficient,
  title={Efficient quantum amplitude encoding of polynomial functions},
  author={Gonzalez-Conde, Javier and Watts, Thomas W and Rodriguez-Grasa, Pablo and Sanz, Mikel},
  journal={Quantum},
  volume={8},
  pages={1297},
  year={2024},
  publisher={Verein zur F{\"o}rderung des Open Access Publizierens in den Quantenwissenschaften}
}

@article{zylberman2024efficient,
  title={Efficient quantum state preparation with Walsh series},
  author={Zylberman, Julien and Debbasch, Fabrice},
  journal={Physical Review A},
  volume={109},
  number={4},
  pages={042401},
  year={2024},
  publisher={APS}
}

@book{nielsen2010quantum,
   title =     {Quantum computation and quantum information},
   author = {Michael A. Nielsen and Isaac L. Chuang},
   publisher = {Cambridge University Press},
   isbn =      {0521635039; 9780521635035; 0521632358; 9780521632355},
   year={2010},
   series =    {Cambridge Series on Information and the Natural Sciences},
   edition = {10th Anniversary},
   pages={555-557}
}

@article{mottonen2004quantum,
  title={Quantum circuits for general multiqubit gates},
  author={M{\"o}tt{\"o}nen, Mikko and Vartiainen, Juha J and Bergholm, Ville and Salomaa, Martti M},
  journal={Physical review letters},
  volume={93},
  number={13},
  pages={130502},
  year={2004},
  publisher={APS}
}

@article{farias2025quantum,
  title={Quantum encoder for fixed-Hamming-weight subspaces},
  author={Farias, Renato MS and Maciel, Thiago O and Camilo, Giancarlo and Lin, Ruge and Ramos-Calderer, Sergi and Aolita, Leandro},
  journal={Physical Review Applied},
  volume={23},
  number={4},
  pages={044014},
  year={2025},
  publisher={APS}
}

@article{melnikov2023quantum,
  title={Quantum state preparation using tensor networks},
  author={Melnikov, Ar A and Termanova, Alena A and Dolgov, Sergey V and Neukart, Florian and Perelshtein, MR},
  journal={Quantum Science and Technology},
  volume={8},
  number={3},
  pages={035027},
  year={2023},
  publisher={IOP Publishing}
}

@article{iaconis2024quantum,
  title={Quantum state preparation of normal distributions using matrix product states},
  author={Iaconis, Jason and Johri, Sonika and Zhu, Elton Yechao},
  journal={npj Quantum Information},
  volume={10},
  number={1},
  pages={15},
  year={2024},
  publisher={Nature Publishing Group UK London}
}

@article{manabe2025state,
  title={The state preparation of multivariate normal distributions using tree tensor network},
  author={Manabe, Hidetaka and Sano, Yuichi},
  journal={Quantum},
  volume={9},
  pages={1755},
  year={2025},
  publisher={Verein zur F{\"o}rderung des Open Access Publizierens in den Quantenwissenschaften}
}

@article{ballarin2025efficient,
  title={Efficient quantum state preparation of multivariate functions using tensor networks},
  author={Ballarin, Marco and Garc{\'\i}a-Ripoll, Juan Jos{\'e} and Hayes, David and Lubasch, Michael},
  journal={arXiv preprint arXiv:2511.15674},
  year={2025}
}

@article{boosari2025hybrid,
  title={Hybrid Quantum State Preparation via Data Compression},
  author={Boosari, Emad Rezaei Fard and Afsary, Maryam},
  journal={arXiv preprint arXiv:2512.01798},
  year={2025}
}

@article{sylvester1867lx,
  title={LX. Thoughts on inverse orthogonal matrices, simultaneous signsuccessions, and tessellated pavements in two or more colours, with applications to Newton's rule, ornamental tile-work, and the theory of numbers},
  author={Sylvester, James Joseph},
  journal={The London, Edinburgh, and Dublin Philosophical Magazine and Journal of Science},
  volume={34},
  number={232},
  pages={461--475},
  year={1867},
  publisher={Taylor \& Francis}
}

@article{arute2019quantum,
  title={Quantum supremacy using a programmable superconducting processor},
  author={Arute, Frank and Arya, Kunal and Babbush, Ryan and Bacon, Dave and Bardin, Joseph C and Barends, Rami and Biswas, Rupak and Boixo, Sergio and Brandao, Fernando GSL and Buell, David A and others},
  journal={nature},
  volume={574},
  number={7779},
  pages={505--510},
  year={2019},
  publisher={Nature Publishing Group UK London}
}

@article{daley2022practical,
  title={Practical quantum advantage in quantum simulation},
  author={Daley, Andrew J and Bloch, Immanuel and Kokail, Christian and Flannigan, Stuart and Pearson, Natalie and Troyer, Matthias and Zoller, Peter},
  journal={Nature},
  volume={607},
  number={7920},
  pages={667--676},
  year={2022},
  publisher={Nature Publishing Group UK London}
}

@article{huang2022quantum,
  title={Quantum advantage in learning from experiments},
  author={Huang, Hsin-Yuan and Broughton, Michael and Cotler, Jordan and Chen, Sitan and Li, Jerry and Mohseni, Masoud and Neven, Hartmut and Babbush, Ryan and Kueng, Richard and Preskill, John and others},
  journal={Science},
  volume={376},
  number={6598},
  pages={1182--1186},
  year={2022},
  publisher={American Association for the Advancement of Science}
}

@article{kim2023evidence,
  title={Evidence for the utility of quantum computing before fault tolerance},
  author={Kim, Youngseok and Eddins, Andrew and Anand, Sajant and Wei, Ken Xuan and Van Den Berg, Ewout and Rosenblatt, Sami and Nayfeh, Hasan and Wu, Yantao and Zaletel, Michael and Temme, Kristan and others},
  journal={Nature},
  volume={618},
  number={7965},
  pages={500--505},
  year={2023},
  publisher={Nature Publishing Group UK London}
}

@article{bluvstein2024logical,
  title={Logical quantum processor based on reconfigurable atom arrays},
  author={Bluvstein, Dolev and Evered, Simon J and Geim, Alexandra A and Li, Sophie H and Zhou, Hengyun and Manovitz, Tom and Ebadi, Sepehr and Cain, Madelyn and Kalinowski, Marcin and Hangleiter, Dominik and others},
  journal={Nature},
  volume={626},
  number={7997},
  pages={58--65},
  year={2024},
  publisher={Nature Publishing Group UK London}
}

@article{google2025quantum,
  title={Quantum error correction below the surface code threshold},
  journal={Nature},
  volume={638},
  number={8052},
  pages={920--926},
  year={2025},
  publisher={Nature Publishing Group UK London}
}

@article{huggins2025efficient,
  title={Efficient state preparation for the quantum simulation of molecules in first quantization},
  author={Huggins, William J and Leimkuhler, Oskar and Stetina, Torin F and Whaley, K Birgitta},
  journal={PRX Quantum},
  volume={6},
  number={2},
  pages={020319},
  year={2025},
  publisher={APS}
}

@article{childs2021high,
  title={High-precision quantum algorithms for partial differential equations},
  author={Childs, Andrew M and Liu, Jin-Peng and Ostrander, Aaron},
  journal={Quantum},
  volume={5},
  pages={574},
  year={2021},
  publisher={Verein zur F{\"o}rderung des Open Access Publizierens in den Quantenwissenschaften}
}

@article{boosari2026hybrid,
  title={Hybrid Quantum Image Preparation via JPEG Compression},
  author={Boosari, Emad Rezaei Fard},
  journal={arXiv preprint arXiv:2602.06201},
  year={2026}
}

@inproceedings{gilyen2019quantum,
  title={Quantum singular value transformation and beyond: exponential improvements for quantum matrix arithmetics},
  author={Gily{\'e}n, Andr{\'a}s and Su, Yuan and Low, Guang Hao and Wiebe, Nathan},
  booktitle={Proceedings of the 51st annual ACM SIGACT symposium on theory of computing},
  pages={193--204},
  year={2019}
}

@article{low2017optimal,
  title={Optimal Hamiltonian simulation by quantum signal processing},
  author={Low, Guang Hao and Chuang, Isaac L},
  journal={Physical review letters},
  volume={118},
  number={1},
  pages={010501},
  year={2017},
  publisher={APS}
}

@book{ahmed2012orthogonal,
  title={Orthogonal transforms for digital signal processing},
  author={Ahmed, Nasir and Rao, Kamisetty Ramamohan},
  year={2012},
  publisher={Springer Science \& Business Media}
}

@article{beauchamp1975walsh,
  title={Walsh functions and their applications.},
  author={Beauchamp, Kenneth George},
  year={1975}
}

@book{sayood2006introduction,
  title={Introduction to data compression},
  author={Sayood, Khalid and others},
  volume={3},
  year={2006},
  publisher={Elsevier}
}
	
\end{document}